# Computational Approach to Dark-Field Optical Diffraction Tomography

*Taean Chang[1,2], Seungwoo Shin[1,2], Moosung Lee[1,2], and YongKeun Park[1,2,3,a]*

[1]*Department of Physics, Korea Advanced Institute of Science and Technology (KAIST), Daejeon 34141, Republic of Korea*
[2]*KAIST Institute for Health Science and Technology, Daejeon 34141, Republic of Korea*
[3]*Tomocube Inc., Daejeon 34051, Republic of Korea*
[a]*To whom correspondence should be addressed. Electronic mail: yk.park@kaist.ac.kr*

The measurement of three-dimensional (3D) images and the analysis of subcellular organelles are crucial for the study of the pathophysiology of cells and tissues. Optical diffraction tomography (ODT) facilitates label-free and quantitative imaging of live cells by reconstructing 3D refractive index (RI) distributions. In many cases, however, the contrast in RI distributions is not strong enough to effectively distinguish subcellular organelles in live cells. To realize label-free and quantitative imaging of subcellular organelles in unlabeled live cells with enhanced contrasts, we present a computational approach using ODT. We demonstrate that the contrast of ODT can be enhanced via spatial high-pass filtering in a 3D spatial frequency domain, and that it yields theoretically equivalent results to physical dark-field illumination. Without changing the optical instruments used in ODT, subcellular organelles in live cells are clearly distinguished by applying a simple but effective computational approach that is validated by comparison with 3D epifluorescence images. We expect that the proposed method will satisfy the demand for label-free organelle observations, and will be extended to fully utilize complex information in 3D RI distributions.

## 1. Introduction

The shapes and dynamics of subcellular structures provide important information about biological cells and tissues. Abnormal shapes of the nucleus membrane and nucleoli are hallmarks of specific types of cancers[1]. The morphology and the dynamics of mitochondria are strongly related to cellular metabolism and can also be altered in response to external environments[2].

Conventionally, subcellular structures are measured using various microscopic techniques. In particular, fluorescent microscopic techniques have been widely used to image subcellular structures, because of their high molecular specificity[3]. Various molecular markers or fluorescent proteins have been developed to label specific subcellular organelles, including MitoTracker$^{TM}$ for mitochondria and Hoechst stains for nucleus[4]. However, fluorescence techniques require exogenous labeling agents such as stains and fluorescent proteins[5] that may cause significant challenges in live cell imaging due to phototoxicity and photobleaching[6]. The expression of the labeling process varies significantly for cells and depends on environmental conditions[7]. The attachment of a fluorescent protein to a target protein may also result in unexpected effects on the function, structure, and localization of the target protein[8].

In contrast, label-free imaging techniques, utilizing intrinsic contrast agents such as vibrational modes[9] or extinction coefficient[10], can ensure non-invasive long-term live-cell imaging. Recent advances in quantitative phase imaging (QPI) techniques exploit refractive index (RI) for label-free and quantitative imaging contrast and have been utilized in studies in various fields[11], including neuroscience[12], biophysics[13], cell biology[14], hematology[15], histopathology[16], and infectious diseases[17]. Among them, optical diffraction tomography (ODT), as a 3-D QPI technique, has been extensively utilized in various research fields[18]. However, images produced by ODT suffer from low molecular specificity, and are difficult to use in the study of subcellular structures. Organelles such as nuclei and lipid droplets can be identified in RI tomograms[19]. However, in many cases, contrasts in RI distributions in cells are not strong enough to clearly distinguish organelles and cellular dynamics.

In this study, we propose a computational method to enhance the imaging contrast in three-dimensional (3D) RI tomography, and experimentally demonstrate its validity. The proposed technique is inspired by the dark-field microscopy that enhances imaging contrasts using a physical filter. The presented method, named Dark-field ODT, performs numerical high-pass filtering in the 3-D spatial frequency domain, and generates contrast-enhanced 3D images of subcellular organelles. We first demonstrate that the contrast enhancement in ODT can be realized via spatial high-pass filtering in 3-D Fourier space, and that it yields theoretically equivalent results to physical dark-field illumination. Without modifying the optical instrument, subcellular organelles in live cells are clearly distinguished, which is validated by comparison with epifluorescence images.

## 2. Principles and Methods

## 2.1. The Transfer Functions of Label-Free Imaging Systems.

Dark-field ODT can be achieved by performing high-pass filtering in the 3D spatial frequency domain. In this section, the principles of Dark-field ODT are explained in terms of a coherent transfer function (CTF). For comparison, the schematics and the optical transfer functions (OTF) of various label-free imaging techniques are shown in Fig. 1.

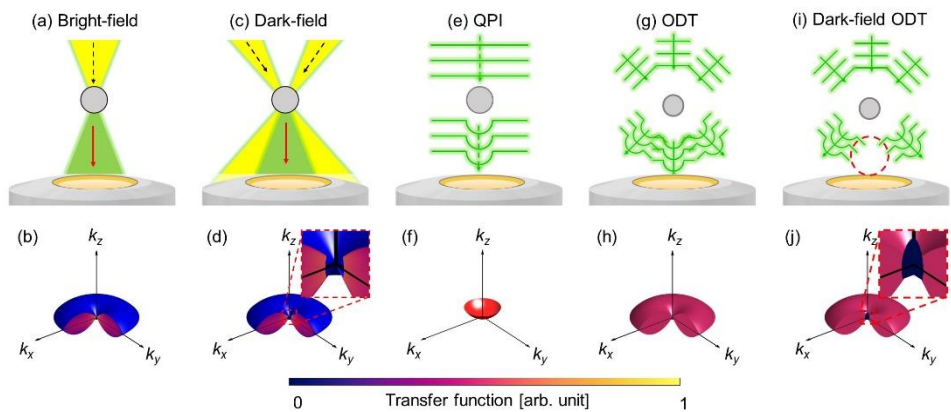

Fig. 1. Schematic illustrations of illuminations and corresponding transfer functions of various techniques. (a, c) Incoherent illumination (yellow) and diffracted signal (green) from a sample in (a) bright-field and (c) dark-field microscopy with (b) and (d) their corresponding 3D OTFs. (e) The QPI schematic and (f) its theoretical 3D CTF. (g, i) ODT and Dark-field ODT with (h, j) their corresponding CTFs. All transfer functions are calculated under the assumption that the numerical apertures of the objective and condenser lenses are the same.

Compared to bright field microscopy [Figs. 1(a) and (b)], dark-field microscopy illuminates a sample with a beam having high spatial frequency components, and the diffracted light from a sample is incoherently summed [Fig. 1(c) and (d)]. Imaging enhancement in dark-field microscopy results from the occlusion of unscattered light, which corresponds to the hole in the origin of an OTF at the spatial frequency domain [the inset of Fig. 1(d)]. By contrast, QPI is a coherent imaging technique that measures the wavefront of a diffracted light[11,20] using interferometry[21], ptychography[22], the transport of intensity equation[23], or the partially coherent transfer function[24] [Fig. 1(e)]. The CTF of the 2D QPI is a portion of a spherical surface in the spatial frequency domain, also known as an Ewald's sphere [Fig. 1(f)]. ODT principles allow reconstructing the 3D RI tomogram of a sample, from multiple 2D holographic images measured at various illumination angles [Fig. 1(g)], and the CTF of ODT can be obtained by accumulating each Ewald's surface corresponding to each 2D holographic image[18] [Fig. 1(h)]. The present Dark-field ODT utilizes the spatial high-pass filtering of 3D RI distributions [Fig. 1(i)], and its CTF is depicted in Fig. 1(j). Dark-field ODT has the same 3D bandwidth as conventional dark-field microscopy in the case of weakly scattering samples, and the Dark-field ODT can even resolve depth information along the optical axis.

## 2.2. Experimental Setup

In order to experimentally demonstrate the proposed method, we exploited ODT systems [Fig. 2]. ODT reconstructs the 3D RI distribution of a weakly scattering sample, from the measured multiple 2D optical fields with various illumination angles, via the principle of inverse scattering[18,25].

The experimental setup is depicted in Fig. 2(a). An off-axis Mach-Zehnder interferometric microscope equipped with a digital micrometer device (DMD) was employed. A coherent laser beam from a diode-pumped solid-state laser (Cobolt Samba™ 532 nm laser) was coupled into a 2×2 fiber coupler. One arm was used for a sample beam, and the other was used as a reference beam. In order to control the illumination angle of the beam impinging a sample, time-multiplexed amplitude holograms displayed on the DMD (DLI 4130, Digital Light Innovations)[26] were projected onto a sample by a condenser lens (UPLSAPO 60XW, Olympus, water 1.2) and a tube lens ($f$ = 300 mm). Then, by an objective lens (PLAPON 60XO, Olympus, oil 1.42) and a tube lens ($f$ = 300 mm), the diffracted beam from a sample was collected and relayed onto an image sensor (LT425-WOCG camera, Lumenera), where its interference pattern with the reference beam was recorded. To further validate the proposed method, a commercial ODT system equipped with 3D fluorescence modality was also used (HT-2H, Tomocube Inc., Republic of Korea) for imaging biological samples (NIH-3T3 cells). The commercial ODT system uses the same wavelength for an illumination beam, and the NA used was 1.2 (water immersion) for both illumination and detection.

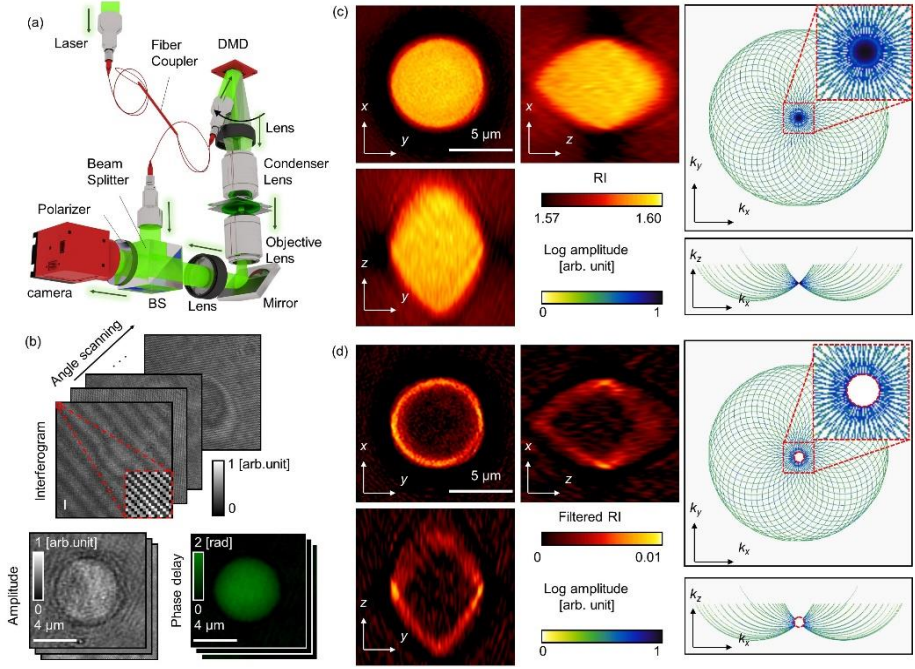

Fig. 2. Principle of Dark-field ODT. (a) An ODT setup equipped with a DMD. (b) Measured interferograms and retrieved optical fields of a polystyrene (PS) microsphere. (c) The RI tomogram of the PS bead (left) and the corresponding Ewald surfaces in the 3D Fourier space (right). (d) Dark-field ODT obtained after applying a high-pass filter to the RI tomogram (left), and the corresponding 3D Fourier space (right).

### 2.3. ODT principles
From the recorded multiple 2D holograms [Fig. 2(b)] of a sample with various illumination angles, both the amplitude and phase images of the sample are retrieved using the off-axis field retrieval algorithm[27]. From these retrieved optical field images of a sample, an RI tomogram is reconstructed by mapping each field to the corresponding Ewald cap in the Fourier space[25] [Fig. 2(c)]. Due to the limited NAs of both the condenser and objective lenses, there are uncollected scattering signals, resulting in the underestimation of RI in ODT, known as the missing cone problem. To address this issue, an iterative algorithm based on the Gerchberg-Papoulis approach with non-negative constraints was used[28]. The number of iterations was eight for all the results obtained in this study. The details on the principles of ODT and the algorithm implemented in MatLab code can be found elsewhere[18,29].

### 2.4. Dark-field ODT algorithm and its physical interpretation
Dark-field ODT is facilitated by applying an additional numerical process of high-pass filtering to the reconstructed RI tomogram [Fig 2(d)]. By blocking out low spatial frequency components in the Fourier space of the reconstructed RI tomogram, it enhances the contrast of the processed 3D RI distribution, and it is a theoretically identical process to conventional dark-field microscopy.

In order to systematically compare the Dark-field ODT to conventional dark-field microscopy, the OTFs of Dark-field ODT are analyzed. Dark-field illumination is the incoherent sum of numerous oblique illuminations slightly beyond the NA of an imaging system. In order to analyze the OTF of dark-field microscopy, a coherent plane-wave illumination slightly beyond the NA is analyzed first. A scattered field $U_s(\mathbf{r}, \mathbf{k}_i)$ under a monochromatic plane-wave illumination $U_i(\mathbf{r}, \mathbf{k}_i) = exp(i\mathbf{k}_i \cdot \mathbf{r})$ contains a subset of sample information (scattering potential $F(\mathbf{r}) = k_0^2[n^2(\mathbf{r})/n_m^2 - 1]/4\pi$). If a scattering potential is slowly varying, the Fourier components of $U_s$ can be directly related to the Fourier components of $F$ within the accuracy of a Rytov approximation[30]:

$$F(\mathbf{K}) = \frac{k_z}{2\pi i} U_i(k_x, k_y) * \psi_s(k_x, k_y) e^{-ik_z z}, \tag{1}$$

where $\mathbf{K} = \mathbf{k} - \mathbf{k}_i$, $k_z = \sqrt{k_0^2 - k_x^2 - k_y^2}$, $\psi_s(\mathbf{r}) = \ln[U_s(\mathbf{r}, \mathbf{k}_i)/U_i(\mathbf{r}, \mathbf{k}_i)]$, $k_0 = 2\pi n_m/\lambda$, $\lambda$ is the wavelength in vacuum, $n_m$ is the RI of a surrounding medium, and $n(\mathbf{r})$ is the 3D RI distribution; the symbols ~ and ∗ denote the Fourier-space representation and convolution operator, respectively. As in Eq. (1), the sample information contained in the scattered field is represented as an Ewald cap in Fourier space.

In dark-field geometry, the radial component of the illumination wavevector $k_{ir}$ is slightly ($\varepsilon \ll 1$) larger than the NA limit of the objective lens $k_{NAr}$, whereas the axial component of the same wavevector $k_{iz}$ is slightly smaller than the NA limit $k_{NAz}$. The DC component of the sample information is not collected because the unscattered light is blocked by an imaging system. The resultant optical fields are related to the Ewald cap of the sample that does not contain the low spatial frequency information.

In the conventional dark-field microscopy, incoherent fields are time-averaged. Theoretically, it can be thought as the sum of coherent fields with randomly varying relative phases. Consequently, the incoherent sum of the fields is identical to the phase-matched coherent sum of the fields[31]. For example, the incoherent sum of scattered optical fields under two plane-wave illuminations results in the mapping of the two Ewald caps corresponding to the illumination angles. A dark-field image is the incoherent sum of scattered fields under numerous circular illuminations at a slightly ($\varepsilon \ll 1$) larger angle than the NA of the optical system ($k_{iz} = k_{NAz} - \varepsilon$). The fields contain information in the $k_z > 0$ region of the accessible Fourier components of the scattering potential, except for the infinitesimal sphere of radius $\varepsilon$ having bandwidth of $-\varepsilon - 2k_{NAr} < k_x < -\varepsilon$, $\varepsilon < k_x < 2k_{NAr} + \varepsilon$, $-\varepsilon - 2k_{NAr} < k_y < -\varepsilon$, $\varepsilon < k_y < 2k_{NAr} + \varepsilon$, and $\varepsilon < k_z < k - k_{NAz} + \varepsilon$. Furthermore, weakly scattering samples have real scattering potentials, which are Hermitian in Fourier space, resulting in information in the $k_z < 0$ region that can be obtained from information in the $k_z > 0$ region using symmetry. Consequently, the bandwidth of dark field microscopy is the same as the bandwidth of Dark-field ODT, as explained in Fig. 1, although the exact transfer function is slightly different due to the difference between coherent and incoherent imaging. As in Fig. 1, the bandwidth is calculated under the assumption that the numerical apertures of the objective and condenser lenses are the same.

*2.5. Filter Selection in Dark-field ODT*

To maximize the enhancement in image contrast while minimizing the artifact due to the addition of high-pass filtering, it is important to choose an appropriate filter and a cut-off frequency. To find an optimal filter type and size, two types of filters (Step and Gaussian) with three different sizes are applied to various simulated and measured RI tomograms (the numerical simulation and experimental measurement of a PS bead, Shepp-Logan phantom, and measured NIH-3T3 cell). The ODT images of these samples are shown in Fig. 3(a).

Spatial frequency responses of the step [Eq. (2)] and Gaussian filter [Eq. (3)] are plotted in Fig. 3(b):

$$H_s\left(\xi_x, \xi_y, \xi_y\right) = 1 \text{ if } |\xi| \geq \xi_c, \ 0 \text{ otherwise.} \tag{2}$$

$$H_G\left(\xi_x, \xi_y, \xi_y\right) = 1 - \exp\left(-\ln 2 \frac{|\xi|^2}{\xi_c^2}\right) = 1 - \left(\frac{1}{2}\right)^{\frac{|\xi|^2}{\xi_c^2}}, \tag{3}$$

where $\xi = k/2\pi$ and the cutoff spatial frequency $\xi_c$ was defined as the frequency of the step rise (step filter) and the frequency of the half response (Gaussian filter), respectively. The filters are isotropic in Fourier space. To reduce the ringing artifacts, a Gaussian filter is chosen from among various types of filters. Figures 4(c)–(h) show the focal plane images of the resultant high-pass filtered RI distributions using the step or Gaussian filters of various $\xi_c$. Ringing artifacts are shown in the case of the step filter, and most of the structures are invisible in the case of filter sizes that are too large [Fig. 3(e)]. Ringing artifacts are significantly reduced in the case of the Gaussian filter, resulting in a clear distinction of boundaries.

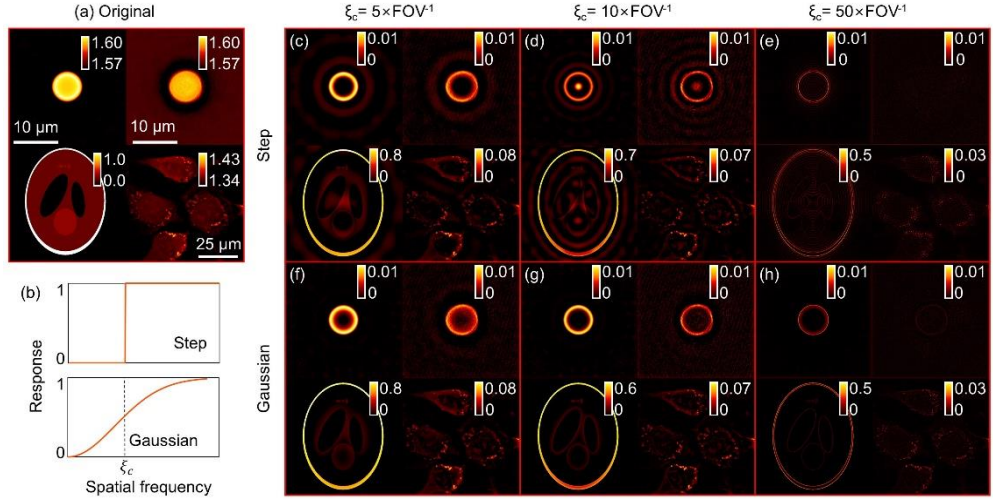

Fig. 3. Implementation of Dark-field ODT with different filter types and sizes. (a) Images of Z slices of various original RI tomograms at the center plane. Results of simulation and experimental measurement of the PS bead (upper left and right images), Shepp-Logan phantom (lower left image), and measured NIH-3T3 cell (lower right image). (b) The spatial frequency response of the step filter and the Gaussian filter of cutoff frequency $\xi_c$. (c, d, e) Images of Z slices of resultant high-pass filtered RI tomograms in (a) using the step filter of $\xi_c = 5 \times \text{FOV}^{-1}$, $\xi_c = 10 \times \text{FOV}^{-1}$, and $\xi_c = 50 \times \text{FOV}^{-1}$. (f, g, h) Images of Z slices of resultant high-pass filtered RI tomograms in (a) using the Gaussian filter of $\xi_c = 5 \times \text{FOV}^{-1}$, $\xi_c = 10 \times \text{FOV}^{-1}$, and $\xi_c = 50 \times \text{FOV}^{-1}$. Ringing artifacts are clearly reduced by using the Gaussian filter.

Because the samples have different scales, cutoff spatial frequencies are expressed in units of an inverse field of view (FOV). The cutoff spatial frequency of the filter should be larger than the spatial frequency (inverse of the size) of the structure to be masked. The beads in Figs. 3(c) and 3(f) are not affected much by the filter of $\xi_c = 5 \times (\text{FOV})^{-1}$ because they are smaller than (FOV)/5. On the other hand, the beads in Figs. 3(d) and 3(g) are affected by the filter of $\xi_c = 10 \times (\text{FOV})^{-1}$ because they are larger than (FOV)/10. The results of the oversized filters are shown in Figs. 3(e) and 3(h), which removes the necessary details altogether. The Gaussian filter with $\xi_c$ slightly larger than the spatial frequency of the structures of interests would be a good choice for most of the samples.

### 2.6. Sample Preparation

As a non-biological sample, 7-μm-diameter polystyrene (PS) microspheres (78462, Sigma Aldrich Inc., USA) in UV curing glue (NOA81, Norland Products Inc., USA) were used. NIH-3T3 cells (ATCC CRL-1658) were used as a biological sample. The cells were maintained in Dulbecco's Modified Eagle's Medium (ATCC 30-2002) supplemented with 10% fetal bovine serum (Invitrogen) at 37°C in a humidified 5% $CO_2$ atmosphere. The cells were confirmed to be free from mycoplasma using an e-Myco Mycoplasma PCR detection kit (iNtRON). The cells in Fig. 6 were plated on a petri dish (Tomodish, Tomocube Inc.) and transfected using Lipofectamine LTX (Invitrogen, #15338-100) according to the manufacturer's instructions. MCherry-conjugated FBL was generated from human brain tissue mRNAs. The cells were stained with Hoechst 33342(0.1 μg/ml, ThermoFisher, Cat. No. H3570). The cells in Fig.6 were also plated on a petri dish (Tomodish, Tomocube Inc.) and stained with MitoTracker Red CMXRos dye (Invitrogen, M7512). The cells in Figs. 5 and 6 were washed with fresh growth medium prior to imaging.

### 3. Results and Discussions

#### 3.1. Simulation and Measurement of Polystyrene Microspheres

In order to validate the proposed method, 7-μm-diameter PS microspheres were measured using ODT and Dark-field ODT, and compared with numerical simulations. Cross-sectional slices of each 3D distribution in the *x–y* and *x–z* planes are presented in Fig. 4. Numerical simulations were performed using a finite-difference time-domain method (FDTD, Lumerical). In the simulation results, the same conditions for the sample and the beam illumination were emulated. RI values of the PS bead and a surrounding medium were set as 1.5983 and 1.574, respectively. The cutoff spatial frequency of the Gaussian filter corresponds to the inverse of the diameter of the bead. As expected, the results from the Dark-field ODT facilitate enhancement in the image contrast and highlight the boundaries of the beads. The simulation and experimental results are consistent, except for the presence of speckle noises in the experimental results that resulted from the use of coherent illumination. The cross-correlation coefficient between them is 0.87. Note that after applying the Dark-field ODT algorithm, the tomogram does not have physical RI values anymore, and it is converted into the filtered RI because the low spatial frequency information is reduced or removed.

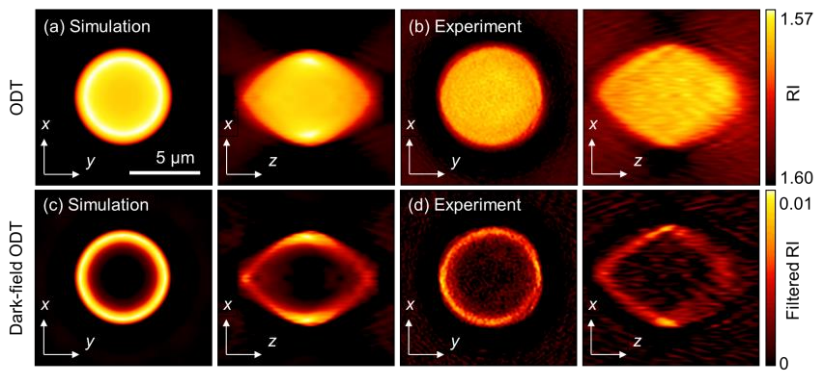

Fig. 4. Simulation and experimental measurements of a 7-μm-diameter PS microsphere. The cross-correlation coefficient between them are 0.87. (a) Cross-sectional slices in the *x*–*y* and *x*–*z* planes of a RI tomogram that is numerically simulated. (b) Cross-sectional slices in the *x*–*y* and *x*–*z* planes of a RI tomogram that is experimentally measured. (c, d) Corresponding Dark-field ODT images retrieved from (a, b). Spatial cutoff frequency corresponds to an inverse of the bead diameter.

### 3.2. Imaging Biological Samples

To demonstrate the applicability to biological samples, a live NIH-3T3 cell was imaged using the proposed method. The epifluorescence signal was also measured in order to confirm the positions of organelles. The nuclei and nucleoli of the cell in Fig. 5 were fluorescently labeled using blue and red fluorescent proteins, respectively. To simultaneously measure both RI and FL in 3D, a commercial ODT setup was used[32] (see Methods).

In Fig. 5, the *x*-*y* cross-sectional slices of the RI tomogram (conventional ODT) at $z = 0$ μm, $z = 0.67$ μm, and $z = 2.33$ μm are presented [Figs. 5(a)–(c)]. Although some distinct organelles with high RI values such as lipid droplets and nuclei can be seen, the image contrasts of these subcellular organelles are not large enough and have similar RI values to that of cytoplasm. Cross-sectional slices of the filtered RI tomogram (Dark-field ODT) in the corresponding planes are presented [Figs. 5(d)–(f)]. The spatial cutoff frequency of the Gaussian filter was set as $(9\ \mu m)^{-1}$ to highlight lower-sized organelles. As a result, the nucleus and nucleoli of the cell become distinguished, whose positions are consistent with the signals in the FL images [Figs. 5(g)–(i)].

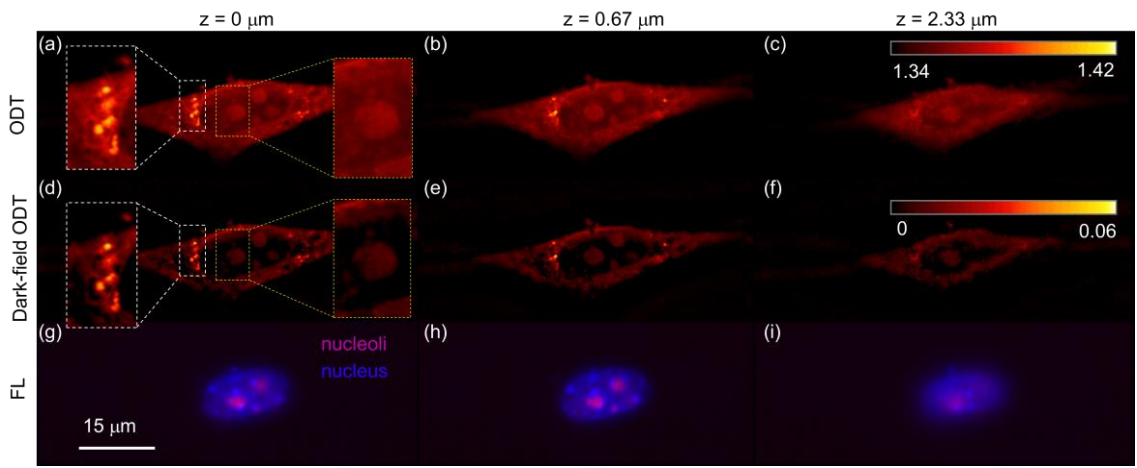

Fig. 5. Application of the Dark-field ODT to a live NIH-3T3 cell for the detection of nucleus and nucleoli. Cross-sectional images of the RI (a–c), filtered RI (d–f), and FL (g–i) tomograms at three different z positions: 0, 0.67, and 2.33 μm. The epi-fluorescence images confirm the positions of the nucleus and nucleoli. The insets show that the nucleoli (the yellow dotted boxes) and the vesicles (the white dotted boxes) are distinguished more clearly in Dark-field ODT.

To further demonstrate the applicability of the present method, we measured highly adherent NIH-3T3 cells, well presenting mitochondria structures. The cross-sectional slice of the RI tomogram (conventional ODT) shows the overall morphology, including cell membrane boundary, nucleus membrane, and vesicles with high RI values [Fig. 6(a)]. However, the RI values of mitochondria and vesicles are similar to that of cytoplasm, resulting in low imaging contrast. This low imaging contrast cannot be enhanced by simply limiting the range of RI values for visualization [Fig. 6(b)]. When the present method is applied to the data [the cutoff spatial frequency of the Gaussian filter was set as (1.5 μm)$^{-1}$], the small organelles including the mitochondria and the vesicles of the cell become more clearly distinguishable [Fig. 6(c)]. The 3D FL image validates that most of the strands in the RI and Filtered RI tomograms are mitochondria. Please note that some of the mitochondria structures in the FL image are not consistent with the filtered RI image, because of the dynamics of mitochondria and the slow acquisition speed of FL. It is reported previously that the correlation between label-free and FL imaging of mitochondria was lower than that of nucleoli[33]. Vesicles with high RI values are regarded as lipid droplets[34], are also clearly distinguishable in Dark-field ODT. Compared to the original RI tomogram. It was reported that most of lipid droplets in cells can be distinguished in RI distribution from its high RI values[34].

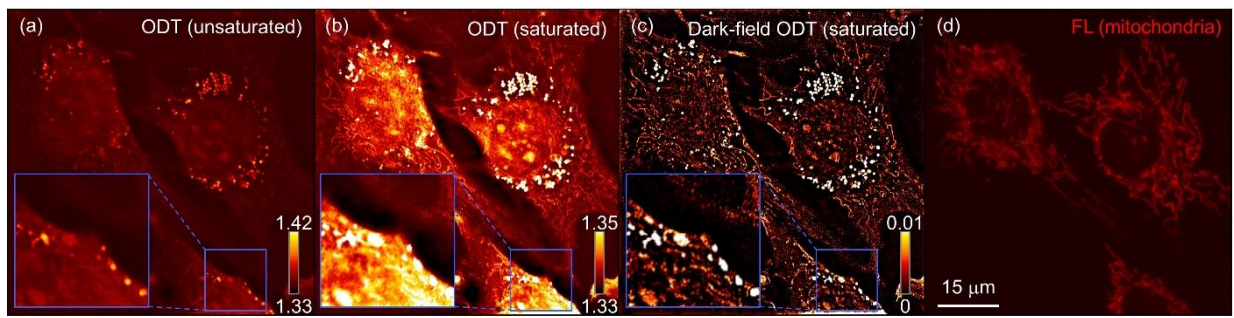

Fig. 6. Application of Dark-field ODT to live NIH-3T3 cells. Cross-sectional images of the RI and filtered RI tomograms in the *x-y* plane are displayed with different colormap ranges. (a) the cross-sectional image of the RI tomogram obtained with the conventional ODT. The colormap ranges containing all RI values in the image (unsaturated colormap range). (b) the cross-sectional image of the RI tomogram with the colormap range highlighting the RI values of mitochondria (saturated colormap range). (c) the cross-sectional image of the filtered RI tomogram obtained with Dark-field ODT. The colormap ranges focusing on the filtered RI values of mitochondria (saturated colormap range).

## 4. Conclusion

The presented Dark-field ODT is a simple but effective computational imaging method to enhance image contrast for 3D QPI. By performing numerical high-pass filtering in a 3D Fourier space, it provides high-contrast imaging for label-free 3D RI tomography. The capability of Dark-field ODT is validated through various types of samples, including numerical simulations of microsphere phantoms and experimental measurements. The Dark-field ODT images of the live NIH-3T3 cells show that small subcellular organelles such as the nucleus membrane, nucleoli, and mitochondria become more clearly distinguishable than they are in the original ODT result. Due to its capability in detecting and imaging small subcellular organelles in live cells, our method can be potentially useful for the study of various biological processes that involve the dynamics and reorganization of organelles.

Because the presented method does not require additional optical setups and the computational burden is very low, it can be readily applied. Moreover, it can provide instantaneous additional imaging modality. Furthermore, the application of the proposed method is not only limited to the field of bioimaging, but can also be expanded to other applications, including the inspections of microlenses or other optical elements[35], the detections of defects in transparent objects[35,36], and the study of materials science.

To fully utilize the presented method in general applications, it is important to select an appropriate filter. Depending on the choice of a filter and its size, the structures of interest could be effectively highlighted; the cutoff spatial frequency should be carefully chosen in order to detail structural information while enhancing imaging contrast. Besides, the application of the process of high pass filtering could result in the formation of image artifacts such as ringing patterns in the case of the step filter. Although two representative filter types and three different filter sizes were demonstrated in this study, a filter can be adjusted and optimized for each application. Considering the recent advances in the combined fields of QPI and machine learning[37], an appropriate choice of a filter can be automatically inferred by training algorithms with multiple image data.

In other imaging fields, the concepts of high-pass filtering of 2D or 3D images have been previously studied. For example, high-pass filtering of the spatial frequency domain has been applied to magnetic resonance imaging and X-ray computed tomography[38]. This work is the first demonstration of the application of frequency filtering in 3D QPI. Furthermore, the application of the high-pass filtering strategy to a 3D Fourier space for coherent imaging is analogous to physical dark-field illumination. It is made possible by the fact that ODT deals with optical field information.

One of the limitations of the proposed method is that the result of Dark-field ODT is a filtered RI, which is not related to the original RI values anymore, because of the application of numerical filtering to spatial frequency information. The value of a filtered RI should be carefully used. Although it can qualitatively provide contrast-enhanced image information, particularly for small objects or sharp edges in a sample, it does not have the quantitative information of RI such as cellular dry mass concentration[39]. Nonetheless, Dark-field ODT only works in the presence of ODT data, therefore, Dark-field ODT can be utilized in correlation with the original ODT. For example, in order to probe the RI value of specific subcellular organelles, the location of organelles can initially be effectively identified using Dark-field ODT. Next, the RI value of the organelles can be retrieved from the original ODT data using the location obtained from the Dark-field ODT. Another limitation is the smaller dynamic range than physical dark-field illumination, which can utilize a full dynamic range of an instrument used. Because insufficient dynamic range can induce the amplified noises, the usability of the method is limited by the signal to noise ratio of the instruments. Nevertheless, as mentioned above, the proposed method does not demand any additional physical setup or heavy computational burden, which means that it is readily accessible for most of ODT users who want to enhance the contrast of their results.

Straightforward applications of the proposed method would include quantitative and correlative studies, exploiting Dark-field ODT with other imaging modalities such as FL[32,40], super-resolution fluorescence imaging[41], Raman microscopy[42], or perturbations of a target sample with thermal excitation[43] or vibrational mode coupling[44]. With the proposed method, we envision that label-free 3D imaging would be further expanded and applied in various fields.

## ACKNOWLEDGMENTS

Mr. Shin and Prof. Park have financial interests in Tomocube Inc., a company that commercializes ODT instruments and is one of the sponsors of the work. This work was supported by KAIST, BK21+ program, Tomocube, and National Research Foundation of Korea (2017M3C1A3013923, 2015R1A3A2066550, 2018K000396). The authors appreciate Dr. Weisun Park (KAIST) and Dr. Sumin Lee (Tomocube Inc.) for providing biological samples used for this study.